\documentclass[%
reprint,
preprintnumbers,
superscriptaddress,
nofootinbib,
 amsmath,amssymb,
 prd,floatfix,
 longbibliography
]{revtex4-1}

\usepackage{graphicx}
\usepackage{dcolumn}
\usepackage{bm}
\usepackage{xcolor}
\usepackage{slashed}
\usepackage{comment}
\usepackage[compat=1.1.0]{tikz-feynman}
\usepackage{hyperref}
\usepackage[capitalize]{cleveref}

\newcommand{\vbb}{$0 \nu \beta \beta $}

\newcommand{\abs}[1]{\left|#1\right|}

\newcommand{\bra}[1]{\ensuremath{\langle #1 |}}   
\newcommand{\ket}[1]{\ensuremath{| #1 \rangle}}   

\newcommand{\amp}[3]{\ensuremath{\langle #1 | #2 | #3 \rangle}} 
\newcommand{\vacuum}[0]{0}
\newcommand{\tsrc}{\ensuremath{t_{\rm src}}}
\newcommand{\tsnk}{\ensuremath{t_{\rm snk}}}
\newcommand{\tfi}{\ensuremath{t^\prime}}
\newcommand{\Pplus}{P_{\texttt{+}}} 
\newcommand{\Z}{\mathbb{Z}}

\usepackage[normalem]{ulem}

\begin{document}

\title{
Long-Distance Nuclear Matrix Elements for  Neutrinoless \\ Double-Beta Decay from Lattice QCD
}

\author{Zohreh~Davoudi}
    \affiliation{Department of Physics and Maryland Center for Fundamental Physics, University of Maryland, College Park, MD 20742, USA}
    \affiliation{Joint Center for Quantum Information and Computer Science, National Institute of Standards and Technology and 
    University of Maryland, College Park, MD 20742, USA}
\author{William~Detmold}
    \affiliation{Center for Theoretical Physics, Massachusetts Institute of Technology, 
    Cambridge, MA 02139, USA}
    \affiliation{The NSF Institute for Artificial Intelligence and Fundamental Interactions, Cambridge, MA 02139, USA}
\author{Zhenghao~Fu}
    \affiliation{Center for Theoretical Physics, Massachusetts Institute of Technology,
    Cambridge, MA 02139, USA}
\author{Anthony~V.~Grebe}
    \affiliation{Center for Theoretical Physics, Massachusetts Institute of Technology, 
    Cambridge, MA 02139, USA}
    \affiliation{Fermi National Accelerator Laboratory, 
    Batavia, IL 60502, USA}
\author{William~Jay}
    \affiliation{Center for Theoretical Physics, Massachusetts Institute of Technology, 
    Cambridge, MA 02139, USA}
\author{David~Murphy}
    \affiliation{Center for Theoretical Physics, Massachusetts Institute of Technology, 
    Cambridge, MA 02139, USA}
\author{Patrick~Oare}
    \affiliation{Center for Theoretical Physics, Massachusetts Institute of Technology, 
    Cambridge, MA 02139, USA}
\author{Phiala~E.~Shanahan}
    \affiliation{Center for Theoretical Physics, Massachusetts Institute of Technology, 
    Cambridge, MA 02139, USA}
    \affiliation{The NSF Institute for Artificial Intelligence and Fundamental Interactions, Cambridge, MA 02139, USA}
\author{Michael~L.~Wagman}
    \affiliation{Fermi National Accelerator Laboratory, 
    Batavia, IL 60502, USA}
\collaboration{NPLQCD Collaboration}

\preprint{FERMILAB-PUB-24-0067-T, MIT-CTP/5682, UMD-PP-024-03}

\date{\today}

\begin{abstract}
Neutrinoless double-beta ($0\nu\beta\beta$) decay is a heretofore unobserved process which, if observed, would imply that neutrinos are Majorana particles.
Interpretations of the stringent experimental constraints on \vbb-decay half-lives require calculations of  nuclear matrix elements. 
This work presents the first lattice quantum-chromodynamics (LQCD) calculation of the matrix element for $0\nu\beta\beta$ decay in a multi-nucleon system, specifically the $nn \rightarrow pp ee$ transition, mediated by a light left-handed Majorana neutrino propagating over nuclear-scale distances.
This calculation is performed with quark masses corresponding to a pion mass of $m_\pi = 806$ MeV at a single lattice spacing and volume.
The statistically cleaner $\Sigma^- \rightarrow \Sigma^+ ee$ transition is also computed in order to investigate various systematic uncertainties.
The prospects for matching the results of LQCD calculations onto a nuclear effective field theory to determine a leading-order low-energy constant relevant for $0\nu\beta\beta$ decay with a light Majorana neutrino are investigated. This work, therefore, sets the stage for future calculations at physical values of the quark masses that, combined with effective field theory and nuclear many-body studies, will provide controlled theoretical inputs to experimental searches of $0\nu\beta\beta$ decay. 

\end{abstract}

\tikz [remember picture, overlay] %
\node [shift={(1cm,-1cm)}] at (current page.north west) %
[anchor=north west] %
{\includegraphics[width=2.5cm]{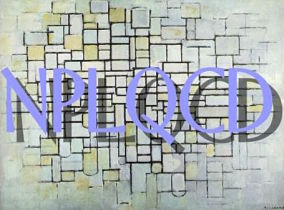}};
                              
\maketitle

\section{Introduction}
\noindent
Neutrinos are the most poorly understood particles within the Standard Model.
In the original conception of the Standard Model, they were presumed to be massless until the discovery of neutrino oscillations~\cite{ray-davis, sno},
which showed that the masses of at least two of the neutrino mass eigenstates are nonzero. The physical mechanism that generates neutrino masses, however, is still uncertain.
If neutrinos are their own antiparticles, their masses could arise through a
Majorana mass term 
\begin{equation}
    -\frac{1}{2} m_{i} \bar{\nu}_{i L} (\nu_{i L})^C + \text{h.c.}
    \label{nu-majorana-mass}
\end{equation}
Here, $(\nu_{iL})^C=C \bar{\nu}_{iL}^T$ with $C$ being the charge-conjugation matrix, and $\nu_{i L}$ is a left-handed neutrino field for each of the mass eigenstates labelled by $i \in \{1,2,3\}$. 
These mass eigenstates are related to the flavor eigenstates $\ell  \in \{ \nu_e,\nu_\mu,\nu_\tau\}$ via $\nu_{\ell L}=\sum_{\ell} U_{\ell, i} \nu_{iL}$, where $U_{\ell,i}$ are the elements of the Pontecorvo–Maki–Nakagawa–Sakata (PMNS) 
mixing matrix~\cite{pontecorvo, mns}.
Alternatively, if the Standard Model is extended to include yet-to-be-observed right-handed neutrinos $\nu_{i R}$, Dirac mass terms $-m_i \bar{\nu}_{i L} \nu_{i R} + \text{h.c.}$ arise naturally, for example, through a Yukawa coupling to the Higgs field analogously to that for the charged leptons.

Resolving whether neutrinos are their own antiparticles, that is, whether terms such as those in Eq.~(\ref{nu-majorana-mass}) are present, is one of the major open problems of modern particle physics.  Since Eq.~(\ref{nu-majorana-mass}) permits lepton-number violation by two units, experimental probes of the Majorana nature of the neutrino search for processes that create and destroy neutrinos in pairs.  

Neutrinoful double-beta ($2\nu\beta\beta$) decay consists of two simultaneous electroweak nuclear transitions in the combined reaction
\begin{equation}
    nn \rightarrow pp ee \bar \nu_{eL} \bar \nu_{eL} \, ,
    \label{neutrinoful-double-beta-equation}
\end{equation}
where two neutrons ($n$) decay into two protons ($p$), two electrons ($e$), and two antineutrinos ($\bar{\nu}_{eL}$).  This process
is the rarest experimentally observed Standard Model process \cite{double-beta-discovery, double-beta-direct-observation}, and only occurs at measurable rates in nuclei that are stable against single-beta decay but favor a double-beta decay. If neutrinos are Majorana, then the two outgoing antineutrinos could mutually annihilate, resulting in a neutrinoless double-beta~(\vbb)-decay
\begin{equation}
    nn \rightarrow pp ee \, ,
    \label{neutrinoless-double-beta-equation}
\end{equation}
which could, in principle, occur in the same nuclei that can undergo $2\nu\beta\beta$ decay. 
Numerous experiments have searched for \vbb~decay \cite{gerda, cuore, KLZ800, double-beta-status-prospects} but, to date, none has conclusively shown that it occurs. 
At present, the most stringent bound on a $0\nu\beta\beta$ decay half-life is $ T^{0\nu\beta\beta}_{1/2} > 2.3\times 10^{26} \text{ yr}$ at 90\% C.L. for $^{136}$Xe from the KamLAND-Zen experiment \cite{KLZ800}.

\begin{figure} [!t] 
    \centering
    \begin{minipage}{0.5\textwidth}
        \centering
        \begin{tikzpicture}
          \begin{feynman}
            \vertex (a1) {\(d\)};
            \vertex [below=2.0 of a1] (b1) {\(d\)}; 
            \vertex [right=6 of a1] (a2) {\(u\)};
            \vertex [below=2.0 of a2] (b2) {\(u\)};
            \vertex [right=3 of a1] (w1);
            \vertex [right=3 of b1] (w2);
            \vertex [below right= 1 of w1] (v1);
            \vertex [above right= 1 of w2] (v2);
            \vertex [right=2 of v1] (el1) {\(e^{-}\)};
            \vertex [right=2 of v2] (el2) {\(e^{-}\)};
            
            \diagram* {
                (a1) -- [fermion] (w1)[crossed dot] -- [fermion] (a2),
                (w1) -- [boson, edge label'=\(W^{-}\)] (v1),
                (v1) -- [plain, edge label'=\(\nu\)] (v2),
                (v1) -- [fermion] (el1),
                (v2) -- [fermion] (el2),
                (w2) -- [boson, edge label=\(W^{-}\)] (v2),
                (b1) -- [fermion] (w2)[crossed dot] -- [fermion] (b2),
            };
          \end{feynman}
        \end{tikzpicture}
    \end{minipage}
\caption{The quark-level diagram responsible for the long-distance contribution to neutrinoless double-beta decay, corresponding to light left-handed Majorana-neutrino exchange between two $W$ bosons.
}
\label{fig:mech}
\end{figure}
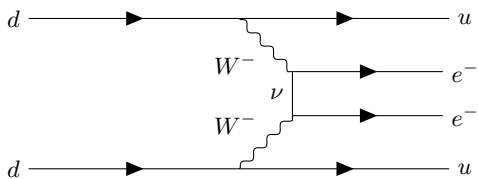

In any theory with a Majorana mass term as in Eq.~(\ref{nu-majorana-mass}), \vbb~decay can be induced via a light left-handed neutrino propagating between two Standard-Model electroweak vertices, as depicted at the quark level in Fig.~\ref{fig:mech}.  Since the left-handed neutrino is nearly massless, the electroweak interactions can be widely separated (up to the diameter of the nucleus undergoing decay), so the resultant interactions are termed long distance.
Beside this minimal extension of the Standard Model, many beyond-the-Standard-Model theories that allow for lepton-number violation,
generate short-distance six-fermion (4-quark--2-electron) effective operators that can also induce \vbb~decay \cite{short-distance-basis, cirigliano-chiral-su3}. The contributions of these operators to the $\pi^- \rightarrow \pi^+ ee$ transition have been studied in Refs.~\cite{callat-short-distance, nplqcd-short-distance} with the lattice-quantum-chromodynamics (LQCD) framework. This work will not consider such short-distance scenarios and focuses on the long-distance mechanism in Fig.~\ref{fig:mech}.

In the light Majorana-neutrino exchange mechanism, the necessity of a helicity flip between the electroweak-current insertions implies that the amplitude for \vbb~decay is proportional to the effective $0\nu\beta\beta$ neutrino mass defined as $m_{\beta\beta} = \left| \sum_{i} U^2_{ei} m_{i}\right|$ and to a hadronic or nuclear matrix element.
A variety of nuclear models have been used to estimate the matrix elements in experimentally relevant nuclei, and significant differences exist between the values predicted by those models \cite{double-beta-status-future, double-beta-status-prospects}.
The resultant model uncertainty can be roughly estimated (but not bounded) by the spread among model predictions and amounts to a factor of three or more. This results in large uncertainties when extracting a bound on $m_{\beta\beta}$ from experimental constraints on half-lives. 
Reducing these uncertainties is crucial for interpreting experimental searches for $0\nu\beta\beta$ decay \cite{Cirigliano:2022oqy, Cirigliano:2022rmf}.

LQCD is a well-established non-perturbative technique for numerically evaluating hadronic and nuclear quantities rooted in quantum chromodynamics (QCD), the theory of the strong force~\cite{ken-wilson, gattringer}.
It, therefore, offers a first-principles method for determining hadronic and nuclear matrix elements relevant to $0\nu\beta\beta$ decay and has been previously used to study $2\nu\beta\beta$ decay \cite{Shanahan:2017bgi, Tiburzi:2017iux}. Nonetheless, the complexity of LQCD computations grows rapidly with baryon number, so initial calculations relevant to \vbb~decay have focused on the mesonic $\pi^- \rightarrow \pi^+ee$ transition \cite{xu-feng-double-beta, detmold-murphy} as a subprocess in a nuclear decay \cite{short-distance-basis, cirigliano-1, cirigliano-2, cirigliano-renormalized}.  This work  extends the approach developed in mesonic calculations to baryonic systems, the $\Sigma^- \rightarrow \Sigma^+ee$ and $nn \rightarrow ppee$ transitions, with the latter relevant to experimental studies in nuclei.
The $nn \rightarrow pp$ transition cannot occur in free space due the the unbound initial state and the dominance of the single-beta decay mode. However, the transition amplitude is well defined and calculable with LQCD even in the absence of a nuclear medium, providing a promising avenue to isolate few-nucleon contributions to the full amplitude in large nuclei. The $\Sigma^-\rightarrow \Sigma^+$ transition also does not correspond to an experimentally observable decay mode (being much slower than the first-order weak decay of the $\Sigma^-$ to $n\pi^-$); it is studied here to
understand systematic uncertainties in the LQCD calculations more thoroughly than can be done with the $nn\rightarrow pp$ transition alone.

By themselves, the LQCD calculations presented here are not sufficient to determine nuclear matrix elements of phenomenological relevance but require connection to nuclear effective field theories (EFTs).
EFTs provide a low-energy description of nuclear processes, including both neutrinoless and neutrinoful double-beta decay, in terms of a set of low-energy constants (LECs) that are \emph{a priori} unknown parameters \cite{nuclear-eft, pionless-eft-1, pionless-eft-2, pionless-eft-3, chiral-eft-1, chiral-eft-2, chiral-eft-3}. 
Matching a (finite-volume) LQCD calculation of the $nn \rightarrow ppee$ transition amplitude to that expressed within a nuclear EFT allows the relevant LECs to be extracted \cite{Cirigliano_2020, cirigliano-master-formula, cirigliano-1, cirigliano-2, zohreh-nu-prop, zohreh-sensitivity}.
Once the systematic uncertainties associated with the present LQCD calculation are fully controlled in future studies, the constrained EFT can be used with many-body methods to calculate nuclear matrix elements in larger nuclei, hence reducing the model uncertainty that currently limits interpretation of experimental results.
The present work explores prospects for the matching procedure to extract a leading-order LEC appearing in the pionless-EFT description of the $nn \to ppee$ process.

\section{Theoretical and Computational Approach}
\label{sec:formalism}
\noindent
This section presents the details of the theoretical and computational approach of this work. After introducing the physical $0\nu\beta\beta$ decay amplitude with a light Majorana neutrino in Sec.~\ref{sec:theory}, Sec.~\ref{sec:correlator}  demonstrates how such an amplitude can in principle be extracted from appropriate two- and four-point correlation functions in LQCD. A more thorough discussion of the exact mapping between the two quantities will be left to Sec.~\ref{sec:matching}.

\subsection{\texorpdfstring{$0\nu\beta\beta$}{0vbb} decay amplitude in the long-range scenario
\label{sec:theory}}
At energies well below the electroweak scale, the Hamiltonian for single-$\beta$ decay is given by 
\begin{equation}
    \mathcal{H}_{W} = 2\sqrt{2} G_{F} V_{ud} (\bar{u}_L \gamma^{\mu} d_L) (\bar{e}_L \gamma_{\mu} \nu_{eL})+\text{h.c.} \, ,
\end{equation}
where $G_F$ is the Fermi constant and $V_{ud}$ is the Cabibbo–Kobayashi–Maskawa (CKM) matrix element encoding the down quark ($d$) to up quark ($u$) transition~\cite{Cabibbo:1963yz,Kobayashi:1973fv}. 
At second order in perturbation theory, this interaction gives rise to a bi-local matrix element of the form
~\cite{Bilenky:2018hbz}
\begin{align}
&\bra{f}S^{(2)}\ket{i} \equiv \frac{(-i)^2}{2!} \int d^{4}x\, d^{4}y \, 
\bra{f} \mathcal{T}\{\mathcal{H}_{W}(x) \mathcal{H}_{W}(y)\} \ket{i} \nonumber\\
    &= - 8 G_F^2 V_{ud}^2 \int d^4 x \, d^4 y \, e^{i(p_1 x + p_2 y)} 
    \mathcal{N}_{e_1} \mathcal{N}_{e_2} \nonumber\\
&\quad\quad \times\bar{u}_{1L}(p_1) \gamma^\mu
\bra{\vacuum} \mathcal{T}\{ \nu_{eL}(x) \nu_{eL}^T(y)\}\ket{\vacuum}
(\gamma^\nu)^T \bar{u}_{2L}^T(p_2) \nonumber\\
&\quad\quad \times \bra{N_f} \mathcal{T}\{ J_\mu(x) J_{\nu}(y)\} \ket{N_i} \, ,
\label{eq:S-realtion-1}
   \end{align}
where $S^{(2)}$ is the second-order contribution to the weak interaction $S$-matrix, and $\bar{u}_{1L}$ and $\bar{u}_{2L}$ are the spinors of the outgoing left-handed electrons with momenta $p_1=(E_1,{\bm p}_1)$ and $p_2=(E_2,{\bm p}_2)$ and state normalization factors $\mathcal{N}_{e_1}$ and $\mathcal{N}_{e_2}$, respectively. 
The quark-level left-handed weak current is 
\begin{equation}
    J_\mu (x) = \bar u_L(x) \gamma_\mu d_L (x),
    \label{eq:weak-current}
\end{equation}
and $\mathcal{T}$ denotes the time-ordering operation. 
The neutrino propagator is given by
\begin{align}
&\bra{\vacuum} \mathcal{T}\{ \nu_{eL}(x) \nu_{eL}^T(y)\}\ket{\vacuum} \nonumber\\
&= -\sum_i U_{ei}^2 m_i \int \frac{d^4 q}{(2\pi)^4} \frac{i}{q^2 - m_i^2 + i \epsilon} e^{-i q\cdot(x-y)} P_L C \nonumber\\
&\approx - m_{\beta\beta} D(x - y) P_L C \, ,
\end{align}
where $P_L=\tfrac{1}{2}(1-\gamma_5)$ is the left-handed projector.
In the last line, the neutrino propagator has factored into a product of Dirac matrices and a massless bosonic propagator,
\begin{align}
    D(x - y) &= \int \frac{d^4 q}{(2\pi)^4} \frac{i }{q^2 + i \epsilon} e^{-i q\cdot(x-y)} \nonumber \\
    &=  \int \frac{d^3 \bm{q}}{(2\pi)^3} \frac{e^{-i |\bm{q}||x^0-y^0| + i\bm{q}\cdot(\bm{x}-\bm{y})}}{2 \abs{\bm{q}}} \, ,
    \label{eq:prop_spectral_rep}
\end{align}
neglecting the neutrino mass compared with momenta characteristic of the hadronic scale. 
Finally, initial and final hadronic states are denoted by $\ket{N_i}$ and $\ket{N_f}$, and are assigned four-momenta $p_{i,f} = (E_{i,f}, \bm{p}_{i,f})$, respectively. Importantly, the spatial momenta of the electrons are set to zero throughout, i.e., $\bm{p}_1 = \bm{p}_2 = 0$.

The $S$-matrix element in Eq.~(\ref{eq:S-realtion-1}) can be simply written as
\begin{align}
\bra{f}S^{(2)}\ket{i} 
=&
 -4 G_F^2 V_{ud}^2 m_{\beta\beta} \mathcal{N}_{e_1} \mathcal{N}_{e_2} \nonumber\\
&\times \int d^4 x \, d^4 y\, L^{\mu\nu}(x, y) H_{\mu\nu}(x, y) \, ,
\label{eq:S-relation-2}
\end{align}
where the hadronic and leptonic tensors are defined via
\begin{align}
H_{\mu\nu}(x, y) &\equiv \bra{N_f} \mathcal{T}\{ J_\mu(x) J_\nu(y)\} \ket{N_i} \, , \label{eq:hadronic_tensor}\\
L^{\mu\nu}(x, y) &\equiv \Gamma^{\mu\nu} D(x- y) \, , \label{eq:leptonic_tensor}
\end{align}
with 
\begin{equation}
\Gamma^{\mu\nu} \equiv \bar{u}_1 \gamma^\mu \gamma^\nu (1 + \gamma_5) C \bar{u}_2^T \, .
\label{eq:Gamma_tensor}
\end{equation}
Here, $\bar{u}_{1,2}$ are the spinors corresponding to the outgoing electrons at rest.

Equation~(\ref{eq:S-relation-2}) can be further processed by inserting a complete set of intermediate hadronic states $\ket{n}$ (with energy $\tilde{E}_n$ and momentum $\tilde{\bm{p}}_n$) between the currents according to $1=\sum_n \tfrac{\ket{n}\bra{n}}{2\tilde{E}_n}$, then using $J_\mu(x)=e^{iP_0x_0-i\bm{P}\cdot \bm{x}}J(0)e^{-iP_0x_0+i\bm{P}\cdot \bm{x}}$ to transform the Heisenberg-picture currents back to the spacetime origin (with $P_0$ and $\bm{P}$ being the energy and momentum operators, respectively). One can then insert the form of the neutrino propagator in Eq.~(\ref{eq:prop_spectral_rep}) and perform integrations over the spacetime coordinates and over the neutrino propagator to arrive at 
\begin{align}
&\bra{f}S^{(2)}\ket{i} = i(2\pi)^4 \delta^4(p_f - p_i +p_1+p_2) \, \mathcal{M}^{i \to f} \, ,
\end{align}
with
\begin{align}
&\mathcal{M}^{i \to f} = 4 G_F^2 V_{ud}^2 m_{\beta\beta} \mathcal{N}_{e_1} \mathcal{N}_{e_2}
    \nonumber \\
    & \times \sum_n 
\left.\frac{\Gamma^{\mu\nu}\big(\bra{N_f}J_\mu(0)\ket{n}\bra{n}J_{\nu}(0)\ket{N_i}+
\mu \leftrightarrow \nu \big)}{4 \tilde{E}_n |\bm{q}|( |\bm{q}| + \tilde{E}_n - E_i + m_e)}\right|_{\bm{q}=\bm{p}_i - \tilde{\bm{p}}_n}.
\label{eq:4-momentum-conservation}
\end{align}
Note that the sum over states $\ket{n}$ involves an implicit integration over the total three-momentum of the intermediate state. 

Considering that the expression in parentheses in the numerator of Eq.~(\ref{eq:4-momentum-conservation}) is symmetric under the exchange of $\mu$ and $\nu$ indices, only the symmetric part of $\Gamma^{\mu\nu}$ contributes to the matrix element. Therefore, one may replace $\gamma^\mu \gamma^\nu$ with $\gamma^{\{\mu} \gamma^{\nu\}} = g^{\mu\nu}$ in Eq.~(\ref{eq:4-momentum-conservation}), giving $\Gamma^{\{\mu\nu\}} = g^{\mu\nu} \Gamma$ with $\Gamma = \bar u_1 (1+\gamma_5) C \bar u_2^T$.  Taking $u_1$ and $u_2$ to have opposite spins (as is required by Pauli exclusion when outgoing momenta vanish), one can show that $\Gamma=1$ up to normalization factors accounted for by $\mathcal{N}_{e_1}$ and $\mathcal{N}_{e_2}$.

Finally, defining the amplitude
\begin{align}
\mathcal{A}^{i \to f} \equiv \frac{\mathcal{M}^{i \to f}}{4G_F^2 V_{ud}^2 m_{\beta\beta}\mathcal{N}_{e_1} \mathcal{N}_{e_2}} \, ,
\label{a-definition}
\end{align}
one obtains
\begin{align}
\mathcal{A}^{i \to f} = \sum_n 
\left.\frac{\bra{N_f}J_\mu(0)\ket{n}\bra{n}J^{\mu}(0)\ket{N_i}}{2 \tilde{E}_n |\bm{q}|( |\bm{q}| + \tilde{E}_n - E_i + m_e)}\right|_{\bm{q}=\bm{p}_i - \tilde{\bm{p}}_n}\,.
\label{a-form}
\end{align}
This quantity encapsulates all of the strong-interaction dynamics of the $0\nu\beta\beta$ decay and is the target of the LQCD calculations discussed below.

\subsection{\texorpdfstring{$0\nu\beta\beta$}{0vbb} decay from LQCD correlation functions
\label{sec:correlator}}
LQCD calculations are performed in Euclidean spacetime to enable Monte Carlo methods. As a result, correlation functions and matrix elements extracted from them are defined in Euclidean spacetime. There are subtleties in the connection between Euclidean and Minkowski matrix elements of time-separated currents when on-shell intermediate states are produced~\cite{Christ:2015pwa,Briceno:2019opb,Davoudi:2020xdv,zohreh-nu-prop}. Nonetheless, as will be discussed later, such states can be avoided in the present calculations; hence Euclidean and Minkowski matrix elements may be related simply by a phase from Wick rotation. As a result, this work will not distinguish Euclidean from Minkowski quantities hereafter but will state the relation between them when necessary. Furthermore, the LQCD study of this work is performed in the isospin limit, corresponding to degenerate up and down quark masses, and does not incorporate electromagnetic interactions. Additionally the electron mass is neglected, $m_e=0$, and consequently $E_i=E_f \equiv E_0$ in the $0\nu\beta\beta$ processes that are studied here. The formalism below is adapted to such a limit. Finally, all quantities are assumed to be defined in an infinite continuous spacetime volume throughout this section. The extension to a discretized finite volume is presented in Secs.~\ref{sec:LQCD} and \ref{sec:matching}.

To proceed, one can define two-point
\begin{align}
C_2(\tfi) =&
\int d^3\bm{x}_f~\bra{\vacuum}\mathcal{O}_i(\bm{x}_f, t_f) 
\mathcal{O}_i^\dagger(\bm{0}, t_i)\ket{\vacuum} \label{eq:c2}
\end{align}

\vspace{40pt}

\noindent and four-point
\begin{align}
C_4^{i\to f}(\tsnk,t,\tsrc) &\equiv
\int d^3\bm{x}_f \, d^3\bm{x} \, d^3\bm{y} \,
D(x-y) \nonumber\\
&\times \bra{\vacuum} \mathcal{O}_{f}(\bm{x}_f, t_f) J_\mu(x) J^{\mu}(y) \mathcal{O}_{i}^\dagger(\bm{0}, t_i) \ket{\vacuum} 
\label{eq:c4}
\end{align}
(Euclidean) correlation functions, which are calculable in LQCD (once spacetime is compactified and discretized).
$\mathcal{O}_i$ and $\mathcal{O}_f$ are  source and sink interpolating operators with the necessary quantum numbers to create the initial and final hadronic states for a given transition.
A similar two-point function to Eq.~(\ref{eq:c2}) can be formed using the final-state interpolating operators but is equivalent to $C_2(t')$ in the isospin limit.
Concrete choices for the interpolating operators will be discussed in Sec.~\ref{sec:LQCD}.
The integrals over the spatial coordinates project the final state and the two currents to zero momentum, so without loss of generality, the source interpolating operator $\mathcal{O}_i^\dagger$ is placed at the spatial origin.
After integrating over spatial coordinates as noted, the correlation functions only depend on the relative (Euclidean) time separations defined as $\tsrc \equiv \min\{t_x, t_y\} - t_i$, $t \equiv t_x - t_y$, $\tsnk \equiv t_f - \max\{t_x, t_y\}$, and $t^\prime \equiv t_f - t_i = t_\text{src} + t_\text{snk} + |t|$, where $t_\text{src}, t_\text{snk} > 0$.

The spectral decomposition of the bi-local matrix element in Eq.~(\ref{eq:c4}) is given by
\begin{widetext}
    \begin{align}
            \bra{\vacuum} &\mathcal{O}_f(\bm{x}_f,t_f) J_\mu(\bm{x},t_x) J^{\mu}(\bm{y},t_y) \mathcal{O}_i^\dagger(\bm{0},t_i) \ket{\vacuum} = \int \frac{d^3\bm{p}_i}{(2\pi)^3} 
            \frac{d^3\bm{p}_f}{(2\pi)^3} 
            \frac{\amp{\vacuum}{\mathcal{O}_f(\bm{x}_f)}{N_f(\bm{p}_f)}\amp{N_i(\bm{p}_i)}{\mathcal{O}_i^\dagger(\bm{0})}{\vacuum}}{2 E_0}
            e^{-E_0 \tfi} \nonumber\\
            &\times \sum_n \frac{\amp{N_f(\bm{p}_f)}{J_\mu(\bm{x})}{n}\amp{n}{J^{\mu}(\bm{y})}{N_i(\bm{p}_i)}}{(2 E_0)(2 \widetilde{E}_n)}  e^{-\Delta \widetilde{E}_{n0} |t|}\left( 1 + A e^{-\Delta E_{10}\tsrc} + B e^{-\Delta E_{10}\tsnk}
            + C e^{-\Delta E_{10}(\tsnk+\tsrc)} + \cdots\right)   \, . 
      \label{eq:c4_spectral_decomp}
    \end{align}
Here, $\Delta E_{n0} \equiv E_n - E_0$ denotes energy splitting between the ground state of the source interpolating operator and the $n$th excited state with the same quantum numbers, while $\Delta\widetilde{E}_{n0} = \tilde{E}_n - E_0$ denotes energy splitting between the source ground state and the $n$th state with the quantum numbers of the intermediate hadronic system. Contributions from backwards-propagating states have been neglected (i.e., an infinite temporal extent is assumed).
The factors $A$, $B$, and $C$ are constants with respect to Euclidean time, expressible in terms of various excited-state matrix elements.
The subleading terms represented by the ellipsis decay at least as quickly as $e^{-\Delta E_{20}\tsrc}$ or $e^{-\Delta E_{20}\tsnk}$.
Similarly, the spectral decomposition of the two-point function takes the form
\begin{align}
\langle \mathcal{O}_i(\bm{x}_f,t_f) \mathcal{O}_i^\dagger(\bm{0},t_i) \rangle
&=  \int \frac{d^3\bm{p}_i}{(2\pi)^3} \frac{\amp{\vacuum}{\mathcal{O}_i(\bm{x}_f)}{N_i(\bm{p}_i)}\amp{N_i(\bm{p}_i)}{\mathcal{O}_i^\dagger(\bm{0})}{\vacuum}}
    {2 E_0} e^{-E_0\tfi}
\Big(1 + D e^{-\Delta E_{10}\tfi} + \cdots\Big) \, ,
\end{align}
where $D$ is constant with respect to Euclidean time. 

The connection to the amplitude in Eq.~\eqref{a-form} is clearest for the ratio of four-point and two-point functions, which can be expressed as
\begin{align}
R^{i \to f}(\tsnk, t, \tsrc)&\equiv
\frac{C_4^{i\to f}(\tsnk,t,\tsrc)}{C_2(\tsnk + |t| + \tsrc)}
\label{eq:ratio_definition}\\
&=
\sum_n 
\left.\frac
    {
    \amp{f}{J_\mu(0)}{n}\amp{n}{J^{\mu}(0)}{i}}
    {(2 E_0)(2\tilde{E} _n)(2\abs{\bm{q}})}
e^{-(\abs{\bm{q}}+\Delta \tilde{E}_{n0}) \abs{t}} \right|_{\bm{q} = -\tilde{\bm{p}}_n} \nonumber\\
&\qquad\times \left(
1 + A e^{-\Delta E_{10}\tsrc} + B e^{-\Delta E_{10}\tsnk}
+ C e^{-\Delta E_{10}(\tsnk+\tsrc)}
- D e^{-\Delta E_{10}\tfi} + \cdots \right) \, .
\label{eq:ratio_spectral_decomp}
\end{align}
\end{widetext}
As indicated, this ratio depends on the three relative operator-time separations.
It then follows that the (Euclidean) amplitude is given by
\begin{align}
\mathcal{A}^{i \to f} &= 2 E_0 \int_{-\infty}^{\infty} dt\, \lim_{\substack{\tsrc \to \infty\\\tsnk \to \infty}} R^{i \to f}(\tsnk, t, \tsrc) \, .
\label{eq:nme_from_ratio}
\end{align}

\section{LQCD Calculation
\label{sec:LQCD}}
\noindent
The LQCD calculation in this work is performed on an ensemble of 12,136 QCD gauge-field configurations separated by 10 trajectories.  The ensemble has a lattice spacing of $a = 0.145$ fm and a volume of $(L/a)^3 \times (T/a) = 32^3 \times 48$. Furthermore, sea quarks are implemented at the $SU(3)$ flavor-symmetric point with degenerate up, down, and strange quark masses corresponding to a pion mass of $m_\pi =806$~MeV. The details of the gauge and fermion actions and the hybrid Monte Carlo scheme used to generate the ensemble are described in Ref.~\cite{nplqcd-1}, with the same action used in other studies of few-baryon systems \cite{nplqcd-1, Beane:2015yha, Chang:2015qxa, Savage:2016mlr, Shanahan:2017bgi, Chang:2017eiq, Davoudi:2019jjk, nplqcd-2, Davoudi:2020ngi, nplqcd-4, Wagman:2021spu, nplqcd-variational, callat-hex-dibaryon}.
Of particular importance for this calculation, the proton, neutron, $\Sigma^0, \Sigma^\pm$, and $\Lambda$ are all degenerate, with a common mass of 
1.64 GeV~\cite{nplqcd-1}.

\subsection{Interpolating operators}

The single-baryon interpolating operators used in this work are
\begin{align}
    \mathcal{O}_{p}^{\sigma} &= [u (\Pplus C\gamma_5) d] (\Pplus u)^\sigma \, ,
    \label{eq:proton-interpolator}
    \\ 
    \mathcal{O}_{n}^{\sigma} &= [d (\Pplus C\gamma_5) u] (\Pplus d)^\sigma \, ,
    \label{eq:neutron-interpolator}
    \\
    \mathcal{O}_{\Sigma^{+}}^{\sigma} &= [u (\Pplus C\gamma_5) s] (\Pplus u)^\sigma \, ,
    \label{eq:sigma-plus-interpolator}
    \\ 
    \mathcal{O}_{\Sigma^{-}}^{\sigma} &= [d (\Pplus C\gamma_5) s] (\Pplus d)^\sigma \, ,
    \label{eq:sigma-minus-interpolator}
\end{align}
where the superscript $\sigma$ is a free spinor index, $C=i\gamma_2\gamma_4$ is the Euclidean charge conjugation matrix, and $\Pplus = (1+\gamma_4) / 2$ is the positive-parity projector.\footnote{The relation between the Minkowski and Euclidean $\gamma$ matrices according to the convention of this work are $\gamma_i^E = -i \gamma_i^M$, $\gamma_4^E = \gamma_0^M$ as given in Ref.~\cite{gattringer}.}
The color and spin contractions implicit in the preceding expressions are defined explicitly for an arbitrary set of three quarks ($q_i\in\{u,d,s\}$) and products of Dirac matrices $(\Gamma_1, \Gamma_2)$ via
\begin{align}
    [q_1 \Gamma_1 q_2] (\Gamma_2 q_3)^\sigma 
    \equiv
    \epsilon_{abc}[(q_1)^{\alpha}_{a} \Gamma_1^{\alpha \beta} (q_2)^{\beta}_{b}] \Gamma_2^{\sigma\delta}  (q_3)^{\delta}_{c} \, ,
\end{align}
where $(\alpha, \beta, \sigma,\delta)$ and $(a,b,c)$ are spin and color indices, respectively, and the square brackets visually isolate the diquark interpolating operator.
The projection of all quarks to positive parity is appropriate for the large quark masses used in the present calculation.
The dinucleon interpolating operators are defined as 
\begin{align}
    \mathcal{O}_{NN} = \mathcal{O}_N^\sigma (C\gamma_5)^{\sigma\sigma^\prime} \mathcal{O}_N^{\sigma^\prime},
\end{align}
where $N\in\{n,p\}$ and the additional $C\gamma_5$ couples the nucleon spins into the required spin-singlet combination.
For the $nn\to pp$ transition, the source and sink operators in the four-point function are $(\mathcal{O}_i,\mathcal{O}_f) = (\mathcal{O}_{nn},\mathcal{O}_{pp})$.
For $\Sigma^-\to\Sigma^+$, the four-point function takes 
$(\mathcal{O}_i, \mathcal{O}_f)=(\mathcal{O}_{\Sigma^-},\mathcal{O}_{\Sigma^+})$.

\subsection{Propagator computation}
\begin{figure}
    \centering
    \includegraphics[width=0.5\textwidth]{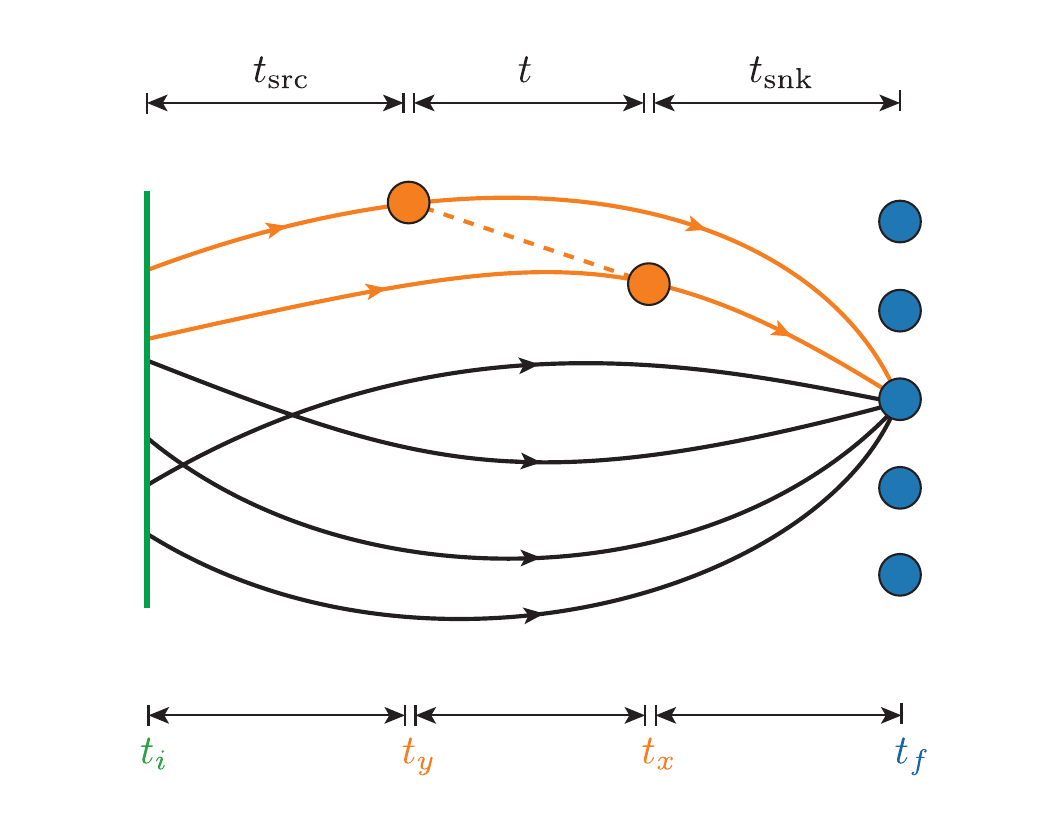}
    \caption{A schematic depiction of the four-point correlation function for the $nn \rightarrow pp$ transition used in this work.  Quark propagators (solid lines) were constructed from a zero-momentum wall source and from point sinks.  Extended propagators, defined in Eq.~(\ref{eq:extended-S}), are denoted by orange lines while the regular, spectator, propagators are shown in black.
    The neutrino propagator (dashed line) between $t_y$ and $t_x$ is given in Eq.~(\ref{eq:neutrino-propagator-continuum}).
    }
    \label{double-beta-diagram}
\end{figure}

The two-point correlation functions in \cref{eq:c2} were computed with a wall source and a point sink.  For the four-point correlation functions in \cref{eq:c4}, propagators were computed originating from both the source and the sink and contracted at the two operator positions $x$ and $y$, as shown in Fig.~\ref{double-beta-diagram}.  
While Eq.~(\ref{eq:c4}) requires summing over all sink interpolating-operator positions, computing propagators from every point at the sink would be prohibitively expensive. Therefore, only a sparse grid of $4^3$ sink points (corresponding to a sparsening factor of $(L/a)/4=8$ in each direction) was used. As studied in Ref.~\cite{nplqcd-sparsening}, this sparse grid corresponds to a partial three-momentum projection and does not modify the low energy spectrum.\footnote{Sparsening was also investigated at the current locations as a means to reduce contraction costs.
However, it was found to produce significant systematic effects on the matrix elements and was ultimately not used.}

On each configuration, spatial grids of point sinks were constructed on every eighth timeslice.
The computationally cheaper zero-momentum wall sources were computed on every timeslice in order to study the effects of varying source-sink separation. A total of 432 propagators were computed on each configuration.

Since all quarks in the interpolating operators in \cref{eq:neutron-interpolator,eq:proton-interpolator,eq:sigma-minus-interpolator,eq:sigma-plus-interpolator} are projected to positive parity, only six (out of twelve) spin-color components of each propagator needed to be computed.
The wall sources with zero three-momentum were constructed in Coulomb gauge with gauge fixing performed in GLU \cite{glu}.
Propagators were computed using the QPhiX inverters \cite{qphix}.\footnote{A minimal wrapper around the underlying inversion functions in QPhiX was developed for this project. Given its general applicability to CPU-based LQCD calculations, the code is made available at \url{https://www.github.com/agrebe/qphix-wrapper}.}

The bosonic propagator associated with the neutrino is defined in a finite periodic Euclidean spacetime in the LQCD calculation. Furthermore, the contribution from the spatial zero momentum is subtracted from the propagator:
\begin{align}
D(x,y) 
=\frac{1}{2L^3}\sum_{\bm{q}\in \frac{2\pi}{L}\Z^3\setminus\{\bm{0}\}}^{|{\bf q}|\leq \pi/a} \frac{1}{\abs{\bm{q}}} e^{i\bm{q}\cdot(\bm{x}-\bm{y})}e^{-\abs{\bm{q}}\abs{t}},
\label{eq:neutrino-propagator-continuum}
\end{align}
where the sum 
runs over non-zero finite-volume momenta and
is truncated at $|\bm{q}| \leq \pi/a$ to regulate the ultraviolet divergence at $x = y$.
This form of the propagator is chosen to make matching to the nuclear EFT seamless~\cite{zohreh-nu-prop}.\footnote{Preliminary studies showed that this form of the propagator also results in less significant short-distance artifacts than the exponentially-regulated form used in a previous study of the $\pi^- \rightarrow \pi^+ ee$ transition in Ref.~\cite{detmold-murphy}.}
The removal of the zero mode ensures that all intermediate states will be at a higher energy than the initial and final states for the volume used in this work, since the minimum neutrino energy is $|\bm{q}|=2\pi/L$.
This approach avoids the difficulties of four-point correlation functions growing exponentially in operator separation times that affected $\pi^- \rightarrow \pi^+ ee$ calculations with a massless intermediate state \cite{xu-feng-double-beta, detmold-murphy}.

\subsection{Contractions}

The four-point correlation function is computationally expensive due to the number of Wick contractions involved and the sums over the sink and both current positions.
First, extended propagators $S^{\alpha\beta, \mu}_{ab}(x)$ were built at the current insertion points $x$ and $y$ via
\begin{align}
    S^{\alpha\beta, \mu}_{ab} (x) = S^{\alpha \delta}_{ae} (x_f|x) \left(J^\mu\right)^{\delta \zeta} S^{\zeta\beta}_{eb} (x|x_i) \, ,
    \label{eq:extended-S}
\end{align}
where $\left(J^\mu\right)^{\delta \zeta}$ is the Dirac structure of the weak current, the propagator $S^{\zeta\beta}_{eb} (x|x_i)$ originates at the source, $S^{\alpha \delta}_{ae} (x_f|x)$ is constructed from the propagator from the sink $S^{\delta\alpha}_{ea} (x|x_f)$ by $\gamma_5$-hermiticity, and dependence on $x_i$ and $x_f$ is left implicit on the left-hand side.
Then, at fixed operator times $t_x, t_y$, two extended propagators were combined with the bosonic propagator $D(x-y)$ (without any spinor or color indices) 
to obtain a four-quark tensor
\begin{align}
    T^{\alpha\beta\gamma\delta}_{abcd} (t_x, t_y) &= \sum_{\bm{x}, \bm{y}} S^{\alpha \beta,\mu}_{ab} (x) D (x-y) 
    S^{\gamma\delta,{\mu}}_{cd}(y) \nonumber\\
    &= \frac{a^3}{L^3} \sum_{\bm{p}} \mathcal{F}[S^{\alpha \beta,\mu}_{ab}](\bm{p}; t_x) \mathcal{F}[D](\bm{p}; t_x - t_y) 
    \nonumber\\
    & \hspace{2.5 cm} \times \mathcal{F}[S^{\gamma\delta,{\mu}}_{cd}](-\bm{p}; t_y) \, ,
    \label{tensor-construction}
\end{align}
with the discrete 3D Fourier transform $\mathcal{F}[f](\bm{p}; t) = \sum_{\bm{x}} e^{i\bm{p} \cdot \bm{x}} f(\bm{x}, t)$ computed efficiently using the fast Fourier transform implemented via the FFTW library \cite{fftw} as in Ref.~\cite{detmold-murphy}.

The tensor in \cref{tensor-construction} was then contracted with the spectator quark propagators connecting the source and sink interpolating operators as prescribed by Wick's theorem to form the four-point $nn \rightarrow pp$ and $\Sigma^- \rightarrow \Sigma^+$ correlation functions.\footnote{
The codebase for the tensor construction and subsequent generation and execution of the Wick contractions for each correlation function can be found at \url{https://www.github.com/agrebe/0vbb}. 
}
The $\Sigma^- \rightarrow \Sigma^+$ correlation function is explicitly given as
\begin{equation}
    \begin{split}
        C_4^{\Sigma^- \rightarrow \Sigma^+} &(t_f, t_x - t_y, t_i) = \epsilon_{abc} \epsilon_{def} \sum_{\bm{x}_f} S_{be}^{\beta\kappa}(x_f | x_i) \\
        &\times \left[ T_{adcf}^{\alpha \delta \sigma \zeta} - T_{cdaf}^{\sigma \delta \alpha \zeta} - T_{afcd}^{\alpha \zeta \sigma \delta} + T_{cfad}^{\sigma \zeta \alpha \delta} \right] (t_x, t_y) \\
        &\times (P_+)^{\sigma \zeta} (P_+ C \gamma_5)^{\alpha \beta} (P_+ C\gamma_5)^{\delta\kappa} \, ,
    \end{split}
\end{equation}
and the $nn \rightarrow pp$ correlation function includes $N_u! N_d! = (4!)^2=576$ terms in the square brackets, each with three additional spectator quark propagators.

Due to the link smearing and improvement in the gauge action and the clover term in the fermion action, time separations of at least three lattice units are required between the current-insertion points and either source or sink locations to avoid contamination from contact terms. 
Subject to this constraint, the four-point correlation function for the $nn \rightarrow pp$ transition was computed at all operator insertions for source-sink separations ranging from 6$a$ to 16$a$, beyond which the statistical noise became prohibitively large.
For the $\Sigma^- \rightarrow \Sigma^+$ transition, where the statistical noise was milder, contractions were computed for all separations less than $T/2=24a$.

\subsection{Renormalization}

The left-handed electroweak-current insertion $J_\mu = \frac{1}{2} \bar u \gamma_\mu (1-\gamma_5) d$ is the difference of vector and axial-vector insertions.
The local lattice currents for these two contributions renormalize separately, so the renormalized current insertion has the form
\begin{equation}
  J_\mu^\text{ren} = \frac{1}{2}\bar u \gamma_\mu (Z_V - Z_A \gamma_5) d \, .
  \label{ew-current-renormalized}
\end{equation}
Due to the interference between the two insertions of these terms in the four-point function, the renormalization factors (or at least the relative renormalization $Z_V/Z_A$) are included at the time the correlation functions are computed.
The renormalization factors for the action parameters used in this work have been computed in Ref.~\cite{lanl-renormalization}:
\begin{align}
Z_V = 0.802(22)\,, && Z_A = 0.879(12)\,.\label{eq:renormalization_factors}
\end{align}

\subsection{Extraction of matrix elements}

\subsubsection{Analysis of two-point functions}

\begin{figure}[ht]
  \centering
  \includegraphics[width=0.50\textwidth]{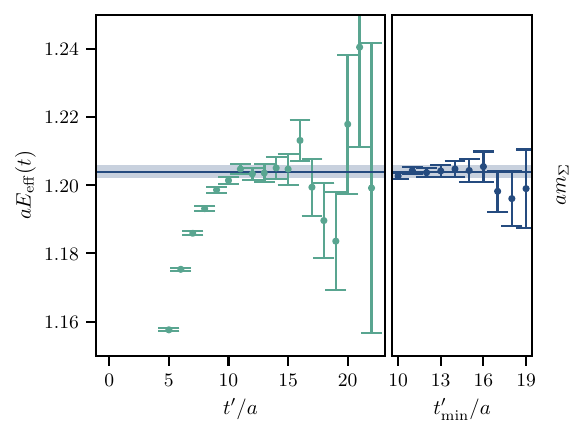}
  \includegraphics[width=0.50\textwidth]{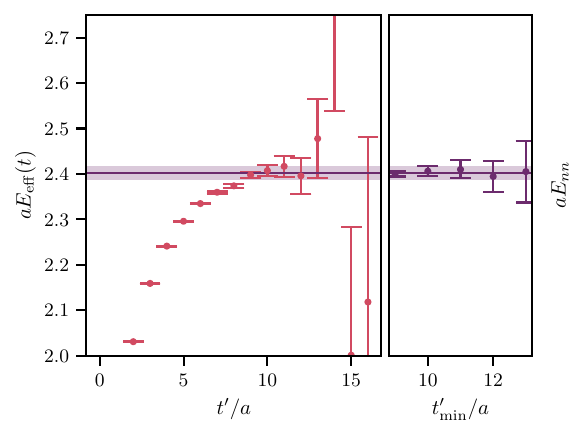}  
  \caption{
    Effective mass (left) and fit results (right) for the $\Sigma$ (upper) and $nn$ (lower) two-point functions.
    The values displayed in the right-hand column are the results of correlated fits to a constant over the temporal extents discussed in the text.
    The horizontal bands show the final results for the masses and the corresponding uncertainties.
    } 
  \label{fig:masses}
\end{figure}

The ground-state energies $m_n=m_\Sigma$ and $E_{nn} = E_{pp}$ are extracted from the respective two-point functions given in \cref{eq:c2}.
\Cref{fig:masses} shows the effective-mass functions for the $\Sigma$ and $nn$ correlation functions, where $aE_{\rm eff}(t')=\ln \left(C_2(t')/C_2(t'+a) \right)$.
Results for fitting the effective mass to a constant using correlated $\chi^2$ minimization are given on the right of \cref{fig:masses} as a function of the minimum time used in the fit.
For $\Sigma$, $t^\prime_{\rm min} \in \{ 10, \dots, 19\}$.
For $nn$, $t^\prime_{\rm min} \in \{9, \dots, 13 \}$ and a cut of $t^\prime_{\rm max} = 16$ is imposed to restrict to points where the statistical noise for the two-point function remains below 30\% of the central value.
Fits with smaller values of $t^\prime_{\rm min}$ were conducted but resulted in poor fit quality ($\chi^2/{\rm dof} > 2$, where ${\rm dof}$ denotes the number of degrees of freedom) and are therefore not shown.
Stability at the level of one standard deviation is observed for the masses extracted from different fits.
The horizontal bands show the result of combined averages and uncertainties using weights based on the Akaike Information Criterion (AIC)~\cite{Jay:2020jkz}.
The final results for the masses in lattice units are
\begin{align}
    a m_\Sigma &= 1.204(2) \, , \label{sigma-mass} \\
    a E_{nn} &= 2.40(2) \, . \label{dineutron-mass}
\end{align}
Note that the interpolating operators used in this work are different from those used in previous studies but yield masses consistent with these earlier studies \cite{nplqcd-1,nplqcd-variational,callat-hex-dibaryon,callat-variational}.
At the level of precision achieved in this study, the dineutron is consistent with either a bound state or a scattering state.

\begin{figure*}[t]
  \centering
  \includegraphics[width=0.825\textwidth]{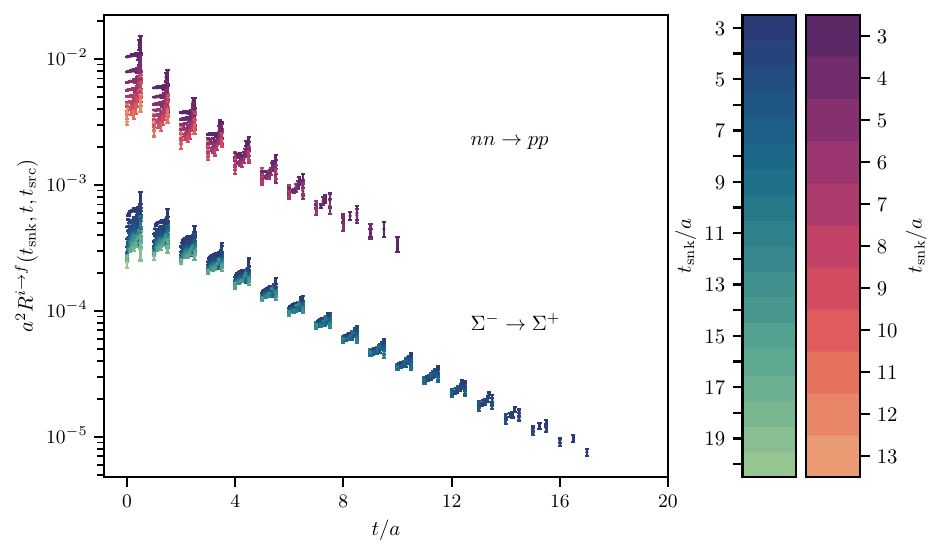}
  \caption{
  The ratio of four-point and two-point correlation functions defined in \cref{eq:ratio_definition} as a function of temporal separation $t$ between the two currents.
  The upper data are for $nn\to pp$, while the lower data are for $\Sigma^- \to \Sigma^+$.
  In both cases, the colors distinguish different values of current-sink separations $\tsnk$ as indicated in the legends.
  For clarity, the points at fixed $(t, \tsnk)$ have been slightly offset in the horizontal direction as $\tsrc$ varies.
  \label{fig:christmas-trees}}
\end{figure*}


\begin{figure*}[t]
  \centering
  \includegraphics[width=0.85\textwidth]{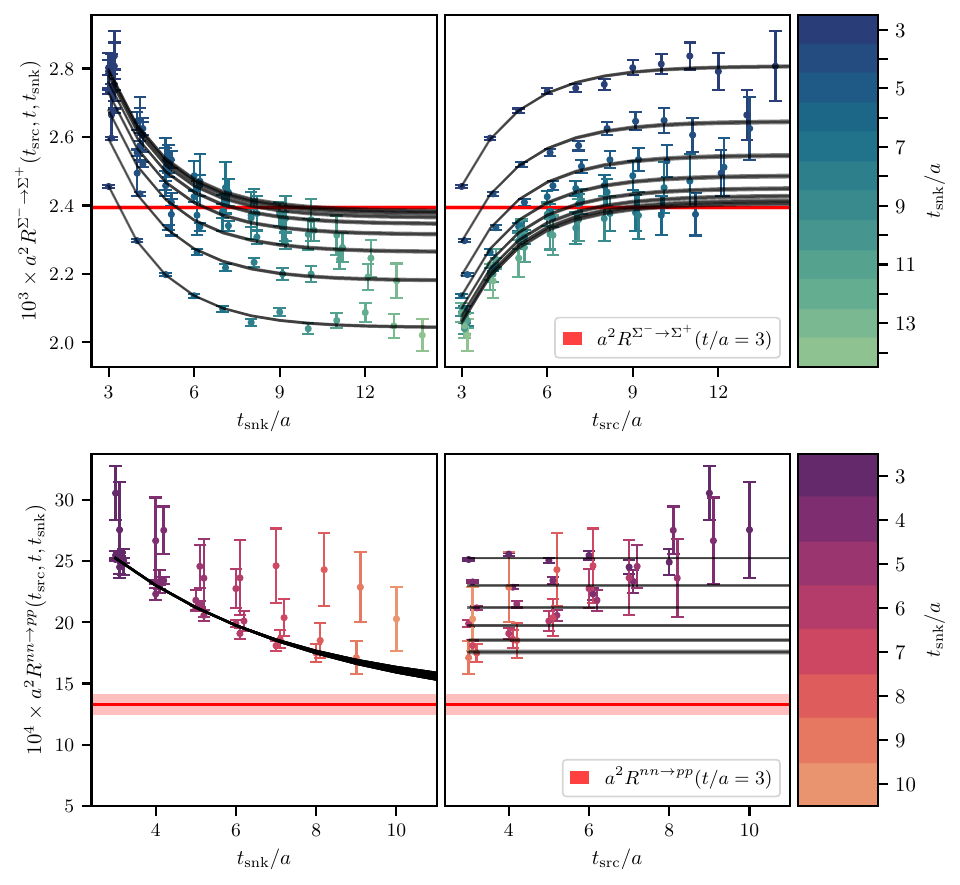}
  \caption{
  The ratio of four-point and two-point correlation functions, $R^{i \to f}(\tsnk, t, \tsrc)$, defined in \cref{eq:ratio_definition}
  for a fixed current separation $t/a=3$.
  The left (right) column shows the dependence on the sink-current (source-current) separation.
  To show the simultaneous dependence on both $\tsnk$ and $\tsrc$, the same data appear in both columns, and matching points appear in the same color on the left and right.
  The solid black curves show the result of a correlated fit to the all the data displayed for a given process ($\Sigma^-\to\Sigma^+$ or $nn \to pp$).
  In each row, the limiting value of $R^{i \to f}(t)$ determined from the fit is shown by the common horizontal line.
  }
\label{fig:joint_fit_fixed_txy}
\end{figure*}

\begin{figure}[!t]
  \centering
  \includegraphics[width=0.48\textwidth]{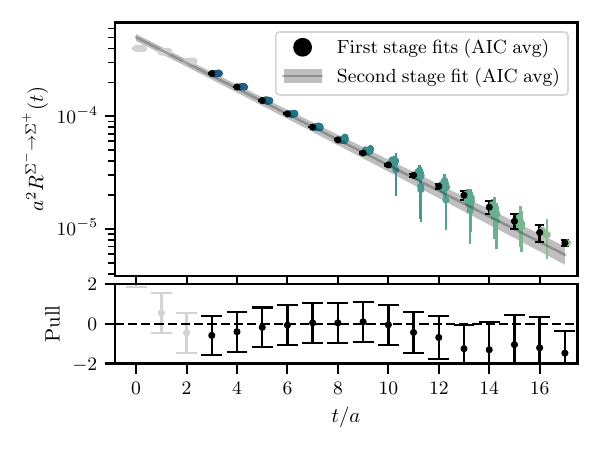}
  \includegraphics[width=0.48\textwidth]{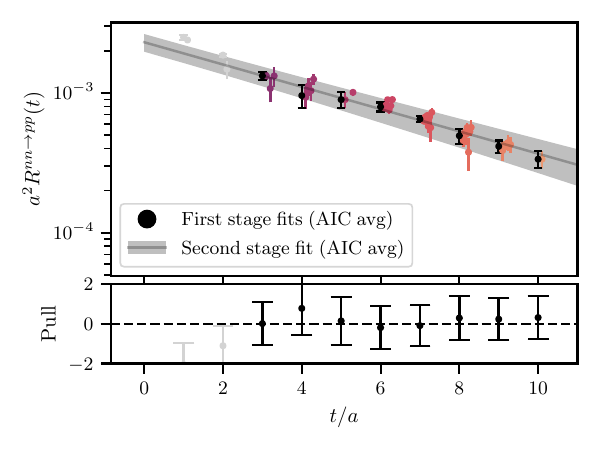}
  \caption{
  The asymptotic ratio $R^{i \to f}(t)$ shown  on a logarithmic scale for $\Sigma^- \to \Sigma^+$ (upper panel) and $nn \to pp$ (lower panel).
  Each cluster of colored points represents fit results at fixed $t$ (with varying $\tsrc$ and $\tsnk$) such as those shown in Fig.~\ref{fig:joint_fit_fixed_txy}.
  The results at each fixed $t$ are combined using model averaging with weights based on the AIC to yield the black points.
  The gray line and error band show the result of second-stage fits to model the dependence on the current-current separation for $t/a\geq 3$.
  The bottom of each panel displays the pull, i.e., the difference between the fit and data in units of the uncertainty. 
  Points excluded from the fit appear in light gray.
  }
  \label{fig:sequential_results}
\end{figure}

\begin{figure}[!t]
  \centering
  \includegraphics[width=0.5\textwidth]{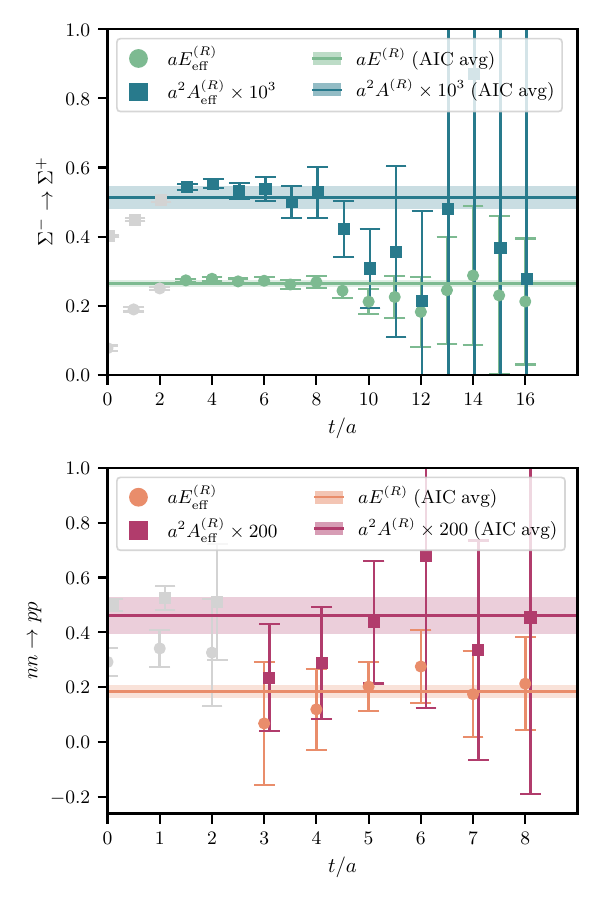}
  \caption{
  Effective energy $E^{(R)}_{\rm eff}$ and amplitude $A^{(R)}_{\rm eff}$ from fits to the ratio $R^{i \to f}(t)$ for $\Sigma^- \to \Sigma^+$ (upper panel) and $nn \to pp$ (lower panel).
  The horizontal lines and error bands show the final posterior results from fits to the exponential decay in \cref{eq:sequential_single_exp}.
  The amplitudes have been re-scaled by arbitrary factors for ease of visualization.
  }
  \label{fig:ratio_meff}
\end{figure}

\subsubsection{Analysis of four-point functions}

The extraction of nuclear matrix elements from a LQCD calculation of the ratio $R(\tsnk,t,\tsrc)$ defined in \cref{eq:ratio_definition} requires controlling excited-state contributions from the source and sink in \cref{eq:ratio_spectral_decomp}, followed by extrapolation and integration over the current separation as in \cref{eq:nme_from_ratio}.
A two-step analysis procedure is used.
First, for fixed current separations, the Euclidean time dependence is modeled with respect to the source and the sink locations to remove excited-state contributions.
The output of the first step is therefore
\begin{align}
    R^{i\to f}(t) \equiv \lim_{\substack{\tsrc \to \infty\\\tsnk \to \infty}} R^{i\to f}(\tsnk, t, \tsrc).
    \label{eq:ratio_limit}
\end{align}
Second, the integral in \cref{eq:nme_from_ratio} must be evaluated to determine the amplitude $\mathcal{A}^{i\rightarrow f}$.
\Cref{eq:ratio_spectral_decomp} shows that $R^{i\to f}(t)$ decays as $\sum_{\bm{q},n}e^{-(\abs{\bm{q}} + \Delta E_{n0})\abs{t}}$.
As shown concretely below, at the present statistical precision and at finite lattice spacing, the sum can be well approximated by a single exponential
\begin{align}
    R^{i\to f}(t) \approx A^{(R)} e^{-E^{(R)} |t|} \, ,
    \label{eq:sequential_single_exp}
\end{align}
where $E^{(R)}$ and $A^{(R)}$ are an effective energy gap and amplitude associated with the asymptotic ratio $R^{i\to f}(t)$.
Departures from this behavior, arising from the full spectrum of states in the sum $\sum_{\bm{q},n}e^{-(\abs{\bm{q}} + \Delta E_{n0})\abs{t}}$ are expected at short times.
However as discussed above, the short-time data ($t/a \leq 2)$ are sensitive to details of the lattice discretization and are excluded from this analysis; subsequent calculations at finer lattice spacings will likely reveal additional contributions to the amplitude from these higher-energy states.
Since these cannot be resolved in the current study, however, the required integral in \cref{eq:nme_from_ratio} can be approximated as
\begin{align}
    \mathcal{A}^{i \to f}
    = 2E_0 \int_{-\infty}^{\infty} dt \, R^{i\to f}(t) \approx  4 E_0 \frac{A^{(R)}}{E^{(R)}} \, .
    \label{eq:nme_2A_by_E}
\end{align}

LQCD results for the ratios $R^{\Sigma^-\to\Sigma^+}(\tsnk, t, \tsrc)$ and $R^{nn \to pp}(\tsnk, t, \tsrc)$ are shown in Fig.~\ref{fig:christmas-trees}, displayed as a function of the temporal separation between the currents.
An alternative view of the data, focusing on the source and sink separations, is given in \cref{fig:joint_fit_fixed_txy} for both $\Sigma^-\to\Sigma^+$ and $nn\to pp$.
As expected from the spectral decomposition, excited-state contamination is generically present from both the source and the sink.
The one exception is for the source-time dependence of $R^{nn\to pp}$, which at the present level of precision is statistically consistent with a constant.\footnote{
    While the fits appear to control excited state contamination well, there is always the possibility of low-lying excited states distorting the results of LQCD calculations, and this concern is of particular importance in the $nn \rightarrow pp$ transition due to the dense low-lying spectrum in nuclear systems \cite{nplqcd-variational}.  Further study with a variety of interpolating operators would be beneficial to confirm the plateau values observed in this work.
}

\textbf{First-stage fits: $\bm{R^{i\to f}(\tsnk, t, \tsrc) \to R^{i \to f}(t)}$.}
For fixed current separation $t$, the data are fit to \cref{eq:ratio_spectral_decomp}.
For $\Sigma^-\to\Sigma^+$, only the leading contributions proportional to $e^{-\Delta E_{10}\tsnk}$ and $e^{-\Delta E_{10}\tsrc}$ are retained (with unknowns $A$, $B$, and $\Delta E_{10}$).
For $nn \to pp$, only the contribution proportional to $e^{-\Delta E_{10}\tsnk}$ is included (with unknowns $B$ and $\Delta E_{10}$), as no dependence on $\tsrc$ is observed within uncertainties.
Examples of the resulting fits are shown by the solid black curves in \cref{fig:joint_fit_fixed_txy}.
The limiting value of $R^{i\to f}(\tsnk, t, \tsrc) \to R^{i \to f}(t)$ emerging from the fit is shown by the common horizontal line.
The fit displayed in the upper row of \cref{fig:joint_fit_fixed_txy} for $\Sigma^-\to \Sigma^+$ has $\chi^2/{\rm dof}$ of $1.04$ for ${\rm dof} = 240$; the fit in the lower row for  $nn\to pp$ has $\chi^2/{\rm dof}$ of $1.15$ for ${\rm dof} = 72$.
Fits of similar quality are obtained for each fixed temporal separation of the currents, yielding $R^{i \to f}(t)$ as a function of $t$.
The results of this process are shown in \cref{fig:sequential_results}.

To verify stability of the fitting procedure, 
the values of $(\tsnk^{\rm  min}, \tsrc^{\rm min})$ included in the fit are varied for each fixed $t$ with $\tsnk^{\rm  min}$ and $\tsrc^{\rm min}$ varied independently in $\{3,4,5,6\}$, which modulates the size of excited-state effects.
To account for any variation in the output values for $R^{i\rightarrow f}(t)$, the results at fixed $t$ are combined using model averaging with AIC weights~\cite{Jay:2020jkz} to yield the black points in \cref{fig:sequential_results}.

\textbf{Second-stage fits: $\bm{R^{i \to f}(t)\to \mathcal{A}^{i \to f}}$.}
As shown in \cref{fig:sequential_results}, $R^{i \to f}(t)$ is saturated by a single decaying exponential for $t/a \geq 3$ in both panels.
This statement is illustrated in \cref{fig:ratio_meff}, which shows the effective energy and effective amplitude
\begin{align}
    aE^{(R)}_{\rm eff}(t) &\equiv \ln \left(\frac{R^{i \to f}(t)}{R^{i \to f}(t+a)}\right)\, , \\
    A^{(R)}_{\rm eff}(t) &\equiv 
    R^{i \to f}(t) e^{E^{(R)}_{\rm eff}(t)\,t} \, .
\end{align}
For $\Sigma^-\to \Sigma^+$, both quantities exhibit  clear plateaus before statistical noise begins to dominate at large times.
For $nn\to pp$, the data are noisier but consistent with a constant.
The data for $R^{i \to f}(t)$ are fit to \cref{eq:sequential_single_exp}, varying $t^{\rm min} \in \{3, 4, \dots, 7 \}$ and $t^{\rm max} \in \{t^{\rm min}+3, t^{\rm min}+4, \dots, t^{\rm  max}_{\rm max}\}$ to check for stability, where the variations in $t^{\rm max}$ extend to $t^{\rm  max}_{\rm max}=10$ for the $nn\rightarrow pp$ transition.  For the $\Sigma^-\to\Sigma^+$ transition, the data were clean enough to allow $t^{\rm  max}_{\rm max}$ to be extended to 17, and a single exponential still sufficed for the second-stage fit.\footnote{The statistically cleaner $\Sigma^-\to\Sigma^+$ channel provides a useful check on the systematic uncertainties of the second stage of the $nn\rightarrow pp$ analysis; fits to $R^{\Sigma^-\to\Sigma^+}(t)$ for $t^{\rm  max}_{\rm max}=10$ are consistent within uncertainties with those with  $t^{\rm  max}_{\rm max}=17$, and consequently, fits with  $t^{\rm  max}_{\rm max}=10$ were also deemed sufficient for the $nn\to pp$ case.}
Results are combined using weights based on the AIC, with the final posterior values for $E^{(R)}$ and $A^{(R)}$ indicated by the horizontal bands in \cref{fig:ratio_meff}.
Due to correlations, the uncertainty in $E^{(R)}$ is somewhat smaller than suggested visually by $E_{\rm eff}^{(R)}$ in \cref{fig:ratio_meff}.
The gray bands in \cref{fig:sequential_results} show the fit results against the data for $R^{i\to f}(t)$. The posterior values for $E^{(R)}$ and $A^{(R)}$ can then be used to evaluate the integral in \cref{eq:nme_2A_by_E}.

The final values for the renormalized amplitudes are
\begin{align}
    a^2 \mathcal{A}^{\Sigma^-\to\Sigma^+} &= 0.00595(58) \, , \label{sigma-final-result} \\
    a^2 \mathcal{A}^{nn \to pp} &= 0.078(16) \, , \label{dineutron-final-result}
\end{align}
where the final uncertainties include both statistical uncertainties and systematic uncertainties from the model averaging as well as the 
uncertainty arising from 
$Z_V$ 
in \cref{eq:renormalization_factors}.
The renormalized amplitude for $\Sigma^-\to\Sigma^+$ is determined with a fractional uncertainty of roughly 10\%, of which the dominant uncertainties are the ratio $A^{(R)}/E^{(R)}$ ($\approx 8\%$) and $Z_V$ ($\approx 5\%$).
The relative breakdown in similar for $nn\to pp$.
The small (few-percent) uncertainty in the ratio $Z_A/Z_V$ is neglected in this work, since it would require recomputing all of the contractions while propagating this uncertainty.

\section{Prospects for Nuclear EFT Matching
\label{sec:matching}}
\noindent
Direct LQCD calculations of $0\nu\beta\beta$ amplitudes in experimentally relevant nuclear isotopes are beyond the reach of the current computational paradigm. The reasons include a substantial increase in complexity of quark-level nuclear correlation functions with increasing atomic number, a severe signal-to-noise degradation of correlation functions as a function of Euclidean time and atomic number, and nuclear excitation gaps that are small compared to the QCD scale which thus demand unrealistically precise spectral resolution.
As a result, nuclear-structure calculations based on nucleonic degrees of freedom, and nuclear-level Hamiltonians and currents, will be the primary method to access phenomenologically relevant nuclear matrix elements for the forseeable future.
These Hamiltonians and currents can be systematically constructed from few-nucleon EFTs, assuming the existence of reliable power-counting schemes.
Nonetheless, such a program is limited by the lack of knowledge of input interactions at the few-nucleon level, particularly for the $0\nu\beta\beta$ process, which has not yet been observed, and importantly, does not occur naturally in few-nucleon systems.
As a result, fully controlled LQCD input at or near the physical values of the quark masses will be crucial in order to constrain unknown low-energy constants (LECs) of the EFTs.

Pionless EFT is a commonly used theoretical framework for studying few-nucleon processes at low energies~\cite{pionless-eft-1, pionless-eft-2, pionless-eft-3, nuclear-eft}. Pionless EFT was applied to the $0\nu\beta\beta$ decay in Refs.~\cite{cirigliano-1,cirigliano-2,cirigliano-renormalized} to determine the amplitude for $nn \to ppee$ process at the lowest EFT orders. Nonetheless, it was found that the EFT amplitude is undetermined for the long-range scenario even at leading order due to the presence of an unknown short-distance LEC, called $g_{\nu}^{NN}(\mu)$, which characterizes the strength of the four-nucleon--two-electron contact interaction at a given renormalization scale, $\mu$. Later studies provided various estimates of this coupling based on a dispersive analysis~\cite{Cirigliano:2020dmx,cirigliano-cottingham} and large-$N_c$ considerations~\cite{Richardson:2021xiu}. However, there remain significant model dependence and uncertainty in these determinations, 
which have been shown to lead to an amplified uncertainty in the nuclear matrix elements in larger nuclear isotopes~\cite{Wirth:2021pij}. Ultimately, LQCD will be able to provide a first-principles determination of this LEC. Such calculations, nonetheless, provide the values of matrix elements in a Euclidean finite spacetime volume, which need to be connected to the physical amplitudes in the corresponding EFT.

Such a formalism for the case of leading-order pionless EFT was developed in Ref.~\cite{zohreh-nu-prop}. Explicitly, the amplitude, defined in Eq.~(\ref{a-definition}), can be related to the leading-order LEC of the EFT by the following matching relation:
\newcommand{\bmom}{{\bar p}}
\begin{widetext}
\begin{equation}
\frac{\mathcal{A}^{nn \to pp} (\bmom_i,\bmom_f)}{2E_0} =
 \bigg [ (1+3g_A^2)
\left(J^{\infty}(\bmom_i,\bmom_f;\mu) + \delta J^V(\bmom_i, \bmom_f) \right) -
\frac{m_n^2}{8\pi^2}\widetilde{g}_\nu^{NN}(\mu) \bigg]
\mathcal{M}(\bmom_i)
\mathcal{M}(\bmom_f) 
\sqrt{\mathcal{R}(\bmom_i) \mathcal{R}(\bmom_f)}
\, .
\label{eq: M^int}
\end{equation}
\end{widetext}
Here, $\bmom_i$ and $\bmom_f$ are the nonrelativistic binding momenta defined as $\bmom_{i,f} = \sqrt{m_n \mathcal{E}_{i,f}}$ for energy shifts $\mathcal{E}_{i,f}$ and the dependence of $\mathcal{A}^{nn\rightarrow pp}$ on these momenta has been made explicit.
$\mathcal{M}(\bmom)$ denotes 
the elastic two-nucleon scattering amplitude in the spin-singlet channel, which can be approximated by an effective-range expansion:
\begin{equation}
    \mathcal{M}(\bmom) = \frac{4\pi}{m_n} \frac{1}{-1/a + r \bmom^2/2 - i\bmom}
\end{equation}
with scattering length $a$ and effective range $r$.
$\widetilde{g}_\nu^{NN}(\mu)$ in Eq.~(\ref{eq: M^int}) is a dimensionless constant related to the LEC $g_\nu^{NN}(\mu)$ by
\begin{align}
\widetilde{g}_\nu^{NN} (\mu) \equiv \left(-\mu + \frac{1}{a}\right)^2 g_\nu^{NN} (\mu),
\end{align}
and $J^\infty(\mu)$ is a known function given by
\begin{align}
J^{\infty}(\bmom_i,\bmom_f;\mu)=\frac{m_n^2}{32\pi^2} \ln\left(\frac{4\pi e^{1-\gamma_E} \mu^2}{-(\bmom_i+\bmom_f)^2-i\epsilon}\right),
\label{eq: Jinf}
\end{align}
with $\gamma_E$ being Euler's constant~\cite{cirigliano-1,cirigliano-2,cirigliano-renormalized}. Furthermore, $\mathcal{R}$ and $\delta J^V$ are two finite-volume functions, whose forms are given in Refs.~\cite{zohreh-nu-prop,zohreh-sensitivity}.  Compared with the matching relation in Eq.~(28) of Ref.~\cite{zohreh-nu-prop} which connects the absolute values of the left and right-hand sides of Eq.~(\ref{eq: M^int}), this work resolves the sign ambiguity in this equation so as to allow for a unique constraint to be placed on the LEC $g_\nu^{NN}(\mu)$.
In the isospin limit where $\bmom_i = \bmom_f \equiv \bmom$, the relation can be simplified as
\begin{align}
\label{eq:simple-matching}
    \frac{\mathcal{A}^{nn \to pp} (\bmom)}{2E_0} &= \bigg[ (1+3g_A^2)
 (J^{\infty}(\bmom,\bmom;\mu) + \delta J^V(\bmom,\bmom) )  \nonumber\\
&\hspace{0.575 cm} -
\frac{m_n^2}{8\pi^2}\widetilde{g}_\nu^{NN}(\mu) \bigg] \left[\mathcal{M}(\bmom)\right]^2 \mathcal{R}(\bmom)
\, .
\end{align}

Despite the relation described in this section, and the LQCD results obtained for $\mathcal{A}^{nn \to pp}(\bmom)$ in this work, several caveats preclude a rigorous determination of $g_\nu^{NN}(\mu)$ via Eq.~(\ref{eq:simple-matching}) at the present time. First and foremost, the LQCD matrix element here is obtained at unphysically large quark masses. Clearly, it is the value of $g_\nu^{NN}(\mu)$ with the physical quark masses that is of phenomenological interest and, \emph{a priori}, the quark-mass dependence of such an LEC is unknown. Therefore, an attempt to constrain $g_\nu^{NN}(\mu)$ or the renormalization-scale--independent quantity
\begin{align}
(1+3g_A^2)
 J^{\infty}(\bmom,\bmom;\mu) -
\frac{m_n^2}{8\pi^2}\widetilde{g}_\nu^{NN}(\mu)
\label{eq:g-scale-indep}
\end{align}
at the quark masses of this work will likely have little bearing on the physical value of the coupling.

Nonetheless, one may still obtain an estimate of the value of this LEC at the quark-mass value of this work, in which case the corresponding values of two-nucleon scattering parameters need to be used in the matching relation.  
To date, there are two classes of LQCD computations of low-energy two-nucleon spectra and scattering parameters at $m_{\pi} \approx 800$ MeV via the use of L\"uscher's finite-volume formalism.  The earlier computations involve asymmetric two-nucleon correlation functions, and point to the existence of rather deep bound states in both the spin-singlet and spin-triplet two-nucleon channels~\cite{nplqcd-1, nplqcd-2, nplqcd-4, callat-hex-dibaryon, yamazaki-1, yamazaki-2}. These were subsequently used to constrain the relevant LECs in electromagnetic and weak reactions of two-nucleon systems at various pion masses and allowed preliminary extrapolations to the physical point~\cite{Beane:2015yha,nplqcd-pp-fusion,Shanahan:2017bgi,Tiburzi:2017iux,Detmold:2021oro}. However, at the finite-volume ground-state two-nucleon energy, which sets the kinematics of the amplitude in this work, the pionless EFT converges poorly when using the values for the effective range and scattering length in those studies. 
Therefore, obtaining the desired $0\nu\beta\beta$-decay amplitude using those results requires extensions of the current leading-order matching formalism, or the use of alternate power-counting schemes. 
The other set of calculations at $m_{\pi} \approx 800$ MeV build symmetric correlation functions to enable accessing the low-lying spectra via a variational method. These lead to upper bounds on ground-state energies that are also consistent with less bound or unbound two-nucleon systems within uncertainties~\cite{mainz-variational, callat-variational, nplqcd-variational}.  No bound states are seen in complementary studies using the Bethe-Salpeter potential method~\cite{halqcd-1, halqcd-2}. While the associated scattering length and effective range for these bounds allow the use of the leading-order matching formalism here, it is non-trivial to turn variational bounds on the energies to bounds on the desired LEC of the EFT, given the nonlinearity of the matching relation.

Despite these caveats, the matching to the EFT amplitude using the above calculation of $\mathcal{A}^{nn \to pp}$, leads to $\tilde g_\nu^{NN}(\mu = m_\pi = 806~\text{MeV})$ values that differ by a factor of four depending on whether the non-variational determinations of two-nucleon energy and scattering parameters or those from the variational studies  are used (assuming the variational bounds are saturated). In both cases, the extracted values are within an order of magnitude of the phenomenological estimate of Ref.~\cite{cirigliano-cottingham}. Consequently, increasingly controlled determinations of the two-nucleon quantities that are input to the matching relation are needed for a robust determination of this LEC. For calculations with physical quark masses, such two-nucleon quantities are well determined phenomenologically, which would ease the matching procedure.

Improving on this situation thus requires 
calculations of ${\cal A}^{nn\to pp}$ and the finite-volume two-nucleon spectrum at or near the physical quark masses.
A point worth emphasizing is that 
the pionless EFT converges at the finite-volume ground-state energy of the spin-singlet two-nucleon system, provided that the lattice volume is sufficiently large, hence putting another requirement on future calculations. For an exploration of the impact of volume on the determination of $g_\nu^{NN}(\mu)$ at the physical values of quark masses, see Ref.~\cite{zohreh-sensitivity}.

\section{Summary and Conclusion
\label{sec:summary}}
\noindent
Within the coming few decades, the sensitivity of experimental neutrinoless double-beta decay searches is projected to increase by several orders of magnitude, corresponding to an order of magnitude decrease in the effective $0\nu\beta\beta$ masses that can be probed~\cite{Cirigliano:2022oqy}. Given current best estimates of nuclear matrix elements, these experiments will likely---but not definitively---be sensitive to the entirety of the parameter space for the inverted hierarchy of neutrino masses.  These searches thus have a large discovery potential but also present the possibility of definitively ruling out the Majorana nature of the neutrino if they find no such decays and if neutrino oscillation experiments confirm the inverted mass hierarchy. Thus, either positive or negative results in next-generation experiments will shed crucial light on this problem provided that the dominant mode of decay is via the exchange of a light Majorana neutrino and that the corresponding nuclear matrix elements can be computed accurately to extract $m_{\beta\beta}$ from measured (bounds on) half-lives.

Starting with the low-energy constants from nuclear effective field theories, nuclear many-body theories can provide \emph{ab initio} calculations of binding energies and $0\nu\beta\beta$ matrix elements in light to moderate ($A \lesssim 48$) nuclei \cite{ab-initio-1, ab-initio-2}. For heavier nuclei ($16 \lesssim A \lesssim 132$), EFT-based approximations to nuclear physics can predict $0\nu\beta\beta$ half-lives with more control than the nuclear models currently used \cite{nuclear-eft-0vbb-1, nuclear-eft-0vbb-2, nuclear-eft-0vbb-3}. As such, determining these low-energy constants in the timescales relevant for these next-generation experiments is of substantial importance to the nuclear- and particle-physics communities \cite{Cirigliano:2022oqy, Cirigliano:2022rmf}.

This work presents 
the first LQCD calculation of the long-distance $0\nu\beta\beta$-decay amplitude of a nuclear system, yielding the result
\begin{equation}
    a^2 \mathcal{A}^{nn \to pp} = 0.078(16)
\end{equation}
on a single LQCD ensemble with a lattice spacing of $a=0.145$ fm, a lattice volume of $(L/a)^3 \times T/a = 32^3 \times 48$, and quark masses corresponding to a pion mass of $m_\pi = 806$ MeV. The baryonic transition ${\cal A}^{\Sigma^-\to\Sigma^+}$ was also determined for the first time.
While this calculation was performed at quark masses that are too large to match to experiment directly, it shows that the relevant matrix elements are calculable in LQCD in multi-baryon systems. This work further discusses prospects for the determination of the leading-order pionless-EFT LEC $g^{NN}_\nu$ from the LQCD matrix element.
Repeating this calculation at lighter quark masses will be non-trivial due to the exponentially worsening signal-to-noise problem as the light-quark masses decrease, a problem especially challenging in multi-baryon systems. However, such calculations are important, as they are the only way to determine experimentally relevant values for the LECs of the nuclear EFTs in a model independent way.

\section{Acknowledgements}

The authors would like to thank Ra\'ul Brice\~no, Balint Jo\'o, Assumpta Parre\~no, Martin Savage, and Andr\'e Walker-Loud for helpful discussions and Marc Illa and Robert Perry for their valuable comments on the manuscript.

Chroma \cite{chroma}, QPhiX \cite{qphix}, and GLU \cite{glu} were instrumental in the calculations of this work, with CPS \cite{cps}, Grid \cite{grid}, and QLua \cite{qlua} playing important roles during code development.
Wolfram Mathematica~\cite{mathematica}, 
numpy~\cite{vanderWalt:2011bqk,Harris:2020xlr}, scipy~\cite{Virtanen:2019joe},
gvar~\cite{gvar}, lsqfit~\cite{lsqfit}, and pandas~\cite{mckinney-proc-scipy-2010,reback2020pandas} were used for data analysis.
Figures were produced using matplotib and seaborn~\cite{Waskom2021,Hunter:2007}.
This work used Stampede2 at the Texas Advanced Computing Center and Anvil at Purdue University through allocation PHY190009 from the Advanced Cyberinfrastructure Coordination Ecosystem: Services \& Support (ACCESS) program, which is supported by National Science Foundation grants \#2138259, \#2138286, \#2138307, \#2137603, and \#2138296 \cite{access}, formerly the Extreme Science and Engineering Discovery Environment (XSEDE), which was supported by National Science Foundation grant number \#1548562 \cite{xsede}. Initial stages of the calculations also made use of the computational resources of the USQCD collaboration.  

ZD was supported by the U.S. Department of Energy, Office of Science, Early Career Award DESC0020271 and by the Department of Physics, Maryland Center for Fundamental Physics, and the College of Computer, Mathematical, and Natural Sciences at the University of Maryland, College Park. WD, ZF, AVG, WJ, DM, PO, and PES were supported in part by the U.S.~Department of Energy, Office of Science, Office of Nuclear Physics, under grant Contract Number DE-SC0011090 and by the U.S. Department of Energy SciDAC5 award DE-SC0023116. WD and PES were also supported by the National Science Foundation under Cooperative Agreement PHY-2019786 (The NSF AI Institute for Artificial Intelligence and Fundamental Interactions, http://iaifi.org/). PES was additionally supported by Early Career Award DE-SC0021006 and by Simons Foundation grant 994314 (Simons Collaboration on Confinement and QCD Strings). AVG and MLW were supported by the resources of the Fermi National Accelerator Laboratory (Fermilab), a U.S. Department of Energy, Office of Science, Office of High Energy Physics HEP User Facility. Fermilab is managed by Fermi Research Alliance, LLC (FRA), acting under Contract No. DE-AC02-07CH11359.


\bibliography{bibi}

\end{document}